\documentclass[12pt]{article}

\usepackage[utf8]{inputenc}      
\usepackage[T1]{fontenc}         
\usepackage{lmodern}             

\usepackage{amsmath,amssymb,amsthm}

\usepackage{graphicx}
\usepackage{siunitx}             

\usepackage[a4paper,margin=1in]{geometry}

\usepackage{float}               
\usepackage{caption}
\usepackage{booktabs}            

\usepackage[%
  backend=biber,
  style=ieee,
  citestyle=numeric-comp,   
  sorting=none,
  doi=false,isbn=false,url=false
]{biblatex}
\usepackage{hyperref}
\hypersetup{
  colorlinks=true,
  linkcolor=blue,
  citecolor=blue,
  urlcolor=blue,
  pdfpagemode=UseNone
}
\usepackage[capitalize,nameinlink]{cleveref}
\crefname{figure}{Fig.}{Figs.}
\crefformat{figure}{#2Fig.~#1#3}

\usepackage{authblk}

\usepackage{todonotes}

\newcommand{\GaAs}{\ensuremath{\mathrm{GaAs} \,}}

\newcommand{\AlGaAs}{\ensuremath{\mathrm{Al}_{0.15}\mathrm{Ga}_{0.85}\mathrm{As} \,}}
\newcommand{\nm}{\nano\metre}
\newcommand{\um}{\micro\metre}

\DeclareUnicodeCharacter{0308}{{\"o}}   
\DeclareUnicodeCharacter{2009}{ }       

\title{Deterministic positioning of circular Bragg gratings using atomic force lithography for high-performance quantum dot light sources}

\author[1]{Sai Abhishikth Dhurjati}
\author[1]{Moritz Langer}
\author[1]{Yared G. Zena}
\author[1]{Ahmad Rahimi}
\author[1]{Liesa Raith}
\author[1]{Martin Bauer}
\author[3]{Frank H. P. Fitzek}
\author[2]{Riccardo Bassoli}
\author[2]{Caspar Hopfmann\thanks{Corresponding author, email: caspar\_arndt.hopfmann@tu-dresden.de}}

\affil[1]{Institute for Emerging Electronic Technologies, IFW Dresden, Helmholtzstraße 20, 01069 Dresden, Germany}
\affil[2]{Quantum Communication Networks research group, Deutsche Telekom Chair of Communication Networks, Dresden University of Technology, Germany}
\affil[3]{Deutsche Telekom Chair of Communication Networks, Dresden University of Technology, Germany}

\begin{document}

\date{\today}

\maketitle

\begin{abstract}
Semiconductor quantum dots (QDs) grown by molecular beam epitaxy are excellent quantum emitters, but their random spatial distribution hinders deterministic coupling to optical microcavities. We demonstrate a room-temperature atomic force microscopy (AFM)–assisted nano-oxidation lithography technique enabling QD positioning with a radial displacement of \qty{51(28)}{\nano\metre}. Free-standing asymmetric circular Bragg gratings incorporating AFM-positioned GaAs QDs exhibit a \num{245}-fold photoluminescence enhancement and fine-structure splitting (FSS) comparable to bulk QDs. Polarization-resolved spectroscopy and finite-difference time-domain simulations show robust emission for displacements up to \qty{50}{\nm} (Stokes parameter $\lvert S \rvert$ $<$ \num{0.05}). The devices display stable FSS and polarization imbalance below \qty{5}{\percent}, confirming precise, reproducible alignment and potential for high fidelity devices. This scalable approach enables deterministic integration of high-performance QDs with photonic cavities, advancing practical quantum light sources for quantum information technologies.

\end{abstract}

\section{Introduction}

The demand for practical quantum technologies and large-scale quantum networks continues to accelerate, placing a growing emphasis on cost-effective and scalable device architectures. At the foundation of these technologies are single-photon and entangled-photon emitters, which are key components for secure quantum communication, photonic quantum computing, and advanced sensing \cite{Claudon2010, Dousse2010, Somaschi2016,Chen2018,Wang2019, Liu2019,Lu2021,Tomm2021,Shooter2020,Uppu2020,Hopfmann2021,Loock2020,Schimpf2021}. Among the various solid-state systems explored, semiconductor quantum dots (QDs) integrated into optical microcavities stand out as particularly promising \cite{Akopian2006, Stevenson2006, Dousse2010}. Their discrete energy levels enable deterministic photon generation with well-defined quantum states, exhibiting Fock state-like photon statistics \cite{Stevens2013, Loock2020}. In addition, QDs offer compatibility with semiconductor processing, wavelength tunability, high emission efficiency via Purcell enhancement, and near-unity photon indistinguishability. Notably, the exciton–biexciton cascade in GaAs QDs provides a highly efficient mechanism for entangled-photon pair emission, positioning these systems as leading candidates for on-chip quantum light sources \cite{Bounouar2015,Huber2017,Chen2018,Wang2019,Hopfmann2021,Langer2025}.

A crucial determinant of the performance of cavity-enhanced quantum light sources is the spatial positioning of the QD relative to the optical cavity mode. Even nanometer-scale misalignment between the emitter and the cavity axis can break in-plane mode symmetry, leading to an imbalanced photon extraction and thereby degradation of polarization entanglement fidelity \cite{Buchinger2025, Peniakov2024}. Despite significant progress in deterministic QD positioning, many device architectures still rely on probabilistic emitter–cavity overlap, which limits scalability and reproducibility.

To address the need for deterministic emitter–cavity alignment, several localization approaches have been developed, including low-temperature micro-photoluminescence ($\mu$PL) mapping \cite{Sapienza2015, Kojima2013, Thon2009, Buchinger2025}, cathodoluminescence imaging \cite{Gschrey2013}, and in situ cryogenic photolithography \cite{Lee2006, Dousse2008}. These techniques have demonstrated excellent positioning precision and have enabled a wide range of cavity–QD architectures. At the same time, their reliance on cryogenic measurements and specialized cryogenic instrumentation motivates the exploration of complementary room-temperature approaches compatible with high-throughput fabrication. AFM-based metrology \cite{Sapienza2017} represents one such route, providing surface localization of QDs relative to lithographic markers. 
Building on this concept, we introduce a room-temperature AFM-based nano-lithography (AFM-NL) approach that integrates AFM surface mapping and nano-oxidation lithography on a single platform, enabling direct, drift-free definition of alignment markers around selected QDs without requiring optical access. Using this method, we fabricate monolithic circular Bragg gratings (CBGs) with deterministically centered QDs and systematically investigate how in-plane displacement influences optical coupling, polarization symmetry, and emission characteristics through Stokes-parameter analysis and finite-difference time-domain (FDTD) simulations. The design and manufacturing process of monolithic CBGs differs from traditional CBGs significantly. Most importantly, instead of using a Gold-back reflector below a transferred semiconductor membrane on top of a buffer layer (e.g. $SiO_2$) as in other high performance CBGs \cite{Liu2019,Wang2019}, monolithic CBGs use vertical asymmetry to achieve comparable vertical emission efficiency values while featuring fewer clean room processing steps. A detailed description and analysis of these devices is found in our recent work \cite{Dhurjati2026a}. Nevertheless, a brief overview of the employed monolithic CBGs in combination with AFM-NL is found in the following.

Our results demonstrate that this approach offers a robust, scalable, and industrially accessible route toward high-performance quantum light sources, bridging the gap between laboratory-scale demonstrations and practical quantum-photonic integration.

\section{Device Concept: Positioned Quantum Dots in Circular Bragg Cavities}
\label{sec:Concept}

\begin{figure}[t!]
    \centering
    \includegraphics[width=0.7\textwidth]{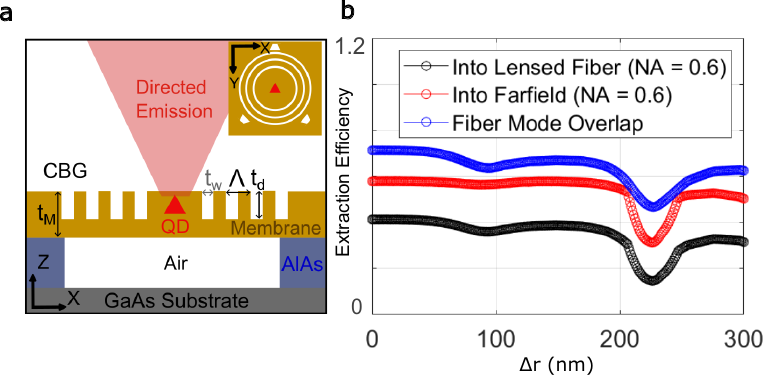}
    \caption{\label{fig:CBG_Concept}
    (a) Schematic illustration of suspended monolithic circular Bragg cavities (CBGs) suitable for AFM-NL positioning around single \GaAs QDs. The principal CBG design parameters are: membrane thickness ($t_M$), trench width ($t_W$), grating period ($\Lambda$), trench depth ($t_d$) and number of CBG rings. (b) FDTD simulation results for a suspended CBG optimized for collection of dipole emission into a lensed single mode fiber with an NA of \num{0.6}. 
    Extraction efficiency, fiber mode overlap, and extraction into lensed single mode fiber as a function of dipole displacement from the CBG center ($\Delta r$). The maximal obtained value of the latter at $\Delta r \to 0$, which is the target value for the CBG parameter optimization process, is \num{0.42}.
 }
\end{figure}

To attain high-performance quantum light sources, and especially high-fidelity entangled photon pair sources, based on \AlGaAs QDs combined with CBGs accurate positioning is essential\cite{Sapienza2015}. This is important not only to maximize the device performance, but also to attain high-fidelity polarization entangled sources. For the latter, broadband emission of a bandwidth on the order of \qty{5}{\nm} is furthermore desirable to efficiently capture photons of both exciton (X) and biexciton (XX) QD transition lines.
As detailed in \cref{sec:AFM-NL}, AFM-NL is a precise method of locating QDs within thin \AlGaAs membranes at room temperature. However, it is highly sensitive to surface contamination. That means that it is very desirable to employ this method before any clean room processing - such as etching, release and transfer of membranes used in traditional CBG fabrication. This fact makes AFM-NL incompatible with the traditional CBG fabrication process, as AFM-NL would need to be employed after the membrane transfer has taken place. Additionally, it would be desirable to avoid the transfer process altogether in order to minimize potentially induced strain in the membrane, which is detrimental, especially with respect to high fidelity entangled photon pair sources \cite{Yang2022}. To mitigate these constraints and enable positioning to single QDs with AFM-NL, suspended monolithic CBGs are used, which feature intentional vertical asymmetry to facilitate efficient vertical light emission. The fabrication and optimization of these micro-cavity structures are described in detail in our recent work \cite{Dhurjati2026a}. A schematic illustration of the suspended monolithic CBGs can be found in \cref{fig:CBG_Concept}a.

For the purpose of this work, the important aspect of this device design approach is that it is fully compatible with AFM-NL positioning of single buried QDs. The device fabrication process can therefore be outlined as follows: First, AFM-NL is used to find and mark single QDs on the \AlGaAs heterostructure surface, \cref{fig:AFM_NL_1}. Next, electron beam lithography is used to align with these markers and to pattern CBG structures on top of these QDs. The monolithic CBGs are consequently fabricated according to the identified target parameters. In this way, high performance quantum light sources are attained.
As seen in Fig. \cref{fig:CBG_Concept}b, the expected emission efficiency into lensed single mode fiber is dependent on the accurate positioning of the QD within the CBG center. According to the finite difference time domain simulations (FDTD), we determine that the relative efficiency stays within \qty{90}{\percent} if the QD is located within a radius of \qty{50}{nm} within the CBG center, see \cref{fig:CBG_Concept}b. A detailed analysis of this effect as a function of emission polarization is found in \cref{sec:Pol_PL}. This demonstrates that accurate positioning to within \qty{50}{nm} is essential to realize highly performant entangled photon pair sources deterministically. Demonstrating this critical aspect is the main premise of this work and is described in detail in the following chapters.

\section{AFM Nano-Oxidation Lithography}
\label{sec:AFM-NL}

\begin{figure}[t!]
    \centering
    \includegraphics[width=0.6\textwidth]{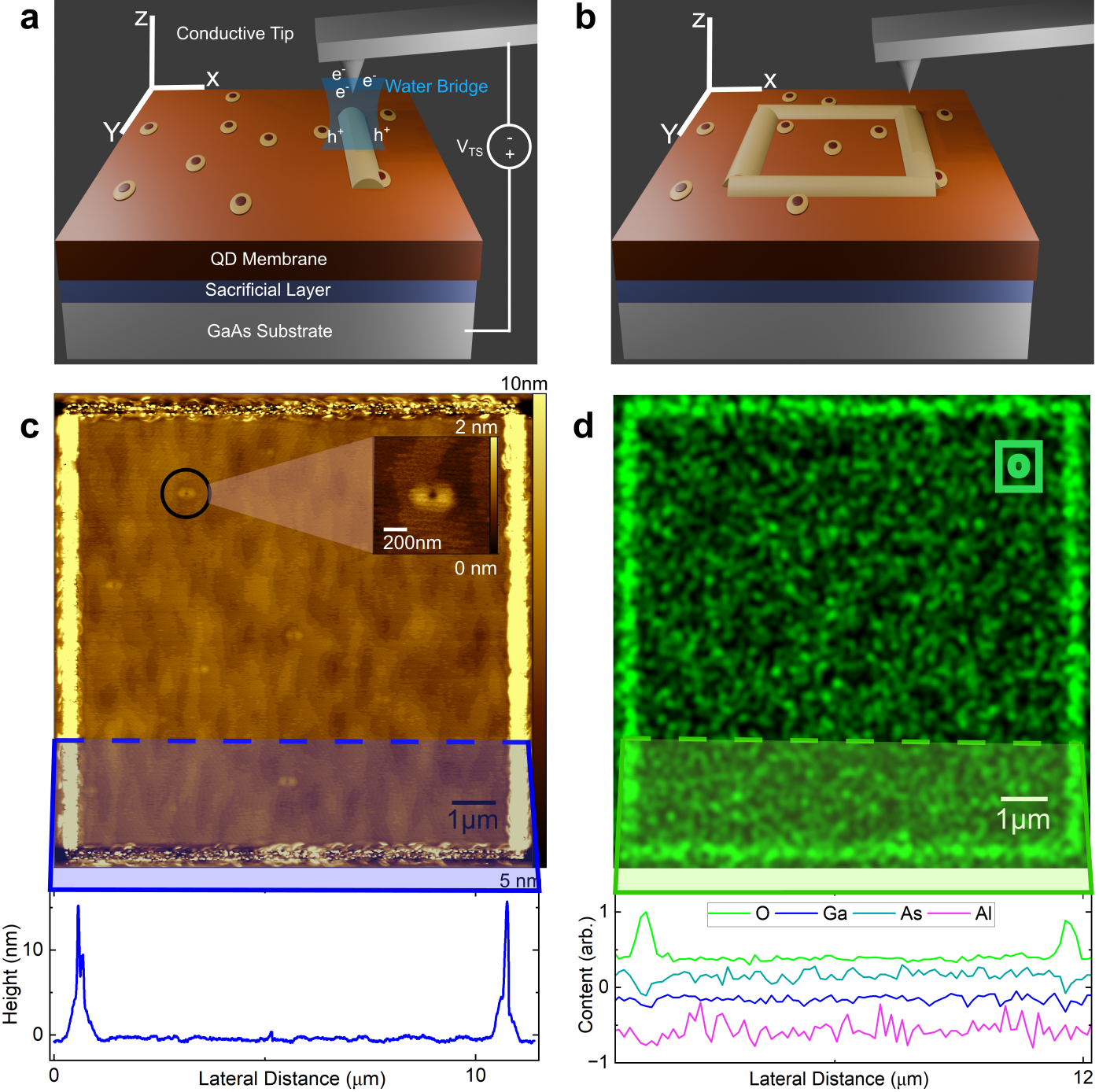}
    \caption{\label{fig:AFM_NL_1}
    (a-b) Schematic overview of the fabrication process for nano-oxidation lithography positioned quantum dots (QDs) for integration into free-standing CBGs.
    (a) AFM-based nano-oxidation lithography of the QD-membrane using a conductive tip.
    (b) Localization of selected QDs followed by the formation of oxide markers. 
    (c) AFM image of a QD-nanomembrane with an oxide marker centered around a single buried QD. Lower panel: AFM height profile across the marker field.
    (d) Energy-dispersive X-ray spectroscopy (EDX) elemental mapping of the oxide marker region, highlighting the spatial oxygen distribution. The bottom panel shows the EDX linescan of the given elements of the processed QD-membrane.  
 }
\end{figure}

$Al_{X}Ga_{1-X}As$ heterostructures are epitaxially grown using molecular beam epitaxy (MBE). GaAs quantum dots (QDs) are formed by infilling droplet etched nanoholes with GaAs and are buried in surrounding Al$_{0.15}$Ga$_{0.85}$As matrix material \cite{Huo2013, Huo2014,Chand1989, Sanguinetti2003, Heyn2007}. Using a GaAs substrate, an AlAs/GaAs superlattice followed by a \qty{250}{\nano\metre} Al$_{0.75}$Ga$_{0.25}$As sacrificial layer and a \qty{150}{\nano\metre} Al$_{0.15}$Ga$_{0.85}$As membrane with centrally embedded QDs are grown.
While these QDs are among the most widely used deterministic and efficient single and entangled-photon pair emitters, their random, self-assembled in-plane growth necessitates precise localization and deterministic positioning for integration into quantum photonic devices. Interestingly, these buried QDs (at a depth of approximately \qty{75}{\nano\metre}) appear as faint but distinguishable nanometer-scale surface features in high-resolution AFM images \cref{fig:AFM_NL_1}c. AFM tips can record sub-nm height signals, enabling sensitive tracing of the heterostructure surface topography, thereby facilitating the identification of individual QDs \cite{Hennessy2007, Huo2013}. Unlike cryogenic optical localization methods that require complex micro-photoluminescence or cathodoluminescence setups, AFM scanning provides a simple and scalable approach to locate buried QDs directly on the sample surface at room temperature. 

The Nanolithography module of the employed AFM (Bruker NanoMan) enables nano-oxidation lithography (AFM-NL), allowing for local surface modification of semiconductor substrates. Previous studies have successfully demonstrated and characterized oxidation writing on GaAs and AlGaAs surfaces by applying a negative voltage (field \qty{>10E9}{\volt \per \metre}) relative to the sample. This results in splitting of $\mathrm{H^{+}}, \mathrm{OH^{-}}$ ions in the water bridge between the tip and sample, creating oxide features on the sample due to the interaction between the $\mathrm{OH^{-}}$ ions and the semiconductor \cite{Cambel2008, Cambel2007, Lin2006}. Building upon this, we exploit AFM-based nanolithography to directly define oxide markers around selected QDs under ambient conditions, without requiring any additional instrumentation. This process employs a conductive PtIr$_5$ coated tip - SCM-PIT; nominal radius \qty{< 25}{\nm}, resonant frequency \qty{\approx 75}{\kilo \hertz}, force constant \qty{\approx 2.8}{\newton \per \metre}. The conductive AFM tip resonates with a mechanical quality factor of approximately \num{250}, with a root-mean-square (RMS) oscillation amplitude of \qty{0.53}{\nano\metre} under typical operating conditions.

During lithography, a bias voltage of $\mathrm{V_{TS}} = $ \qty{-20}{\volt} is applied between the tip and sample, while a water meniscus sustained by a built-in humidifier (\qty{\sim 60}{\percent}) mediates electron transport. This localized electrochemical reaction oxidizes the semiconductor surface along the direction of tip propagation, forming desired GaO$_x$ and AlO$_x$ features \cite{Ahn2011}. Owing to the larger molar volume of the oxide relative to the parent matrix, these regions appear as raised features in AFM height scans \cref{fig:AFM_NL_1}(c) and also provide contrast to energy-dispersive x-ray (EDX) spectroscopy analysis using a scanning electron microscope (SEM). The spatially resolved EDX spectra at kinetic electron energies of \qty{2}{\kilo \volt} show strong oxygen (O) peaks along the marker growth, a constant Aluminum (Al) signal with a minor Gallium (Ga) signal and an absence of Arsenic (As), confirming that oxidation is confined to Al and Ga while As remains in the unmodified lattice \cite{Graham2007}. The attained oxide structures serve as alignment markers for subsequent electron beam lithography (EBL) patterning of CBG microcavities. 

During AFM imaging for QD localization, the cantilever is operated in tapping mode (set point ratio \num{\approx 0.7}) to minimize surface interaction. For oxide writing, the system is switched to contact mode with a deflection set point of \qtyrange{0.1}{0.2}{\volt}; this ensures stable formation of the water meniscus and uniform oxide growth without tip wear. Oxide markers can be written in arbitrary geometries such as periodic arrays or directly patterned photonic structures like circular Bragg gratings or nanoscale quantum devices \cite{Gildemeister2007, Gaikwad2015}. Several previous studies have been performed using nanolithography, including site-controlled InAs/GaAs QDs \cite{MartinSanchez2009, Cha2012, Ohashi2004, Atkinson2009}, current control on 2DEGs \cite{Ishii1995}, and Aharonov–Bohm oscillations in a ring grown in oxide \cite{Keyser2002}. These applications demonstrate the versatility of this approach. Further refinement of AFM positioning technique employed in this work could be achieved by leveraging EBL systems with integrated EDX detectors. Additional datasets and a more detailed discussion of the employed EDX spectroscopy are discussed in the supplementary Sec.2.

Oxide formation in QD-membranes is dependent on the applied bias voltage $\mathrm{V_{TS}}$, Al concentration of the sample, and humidity, providing various controllable parameters, see also \cref{fig:AFM_NL_2}a. At lower voltages, oxidation is limited by the slow drift of ionic species through the nanoscale water bridge, while at higher voltages the electric field drives rapid anodic oxidation, until saturation is typically reached due to space-charge accumulation consistent with the Cabrera–Mott model of field-assisted oxidation \cite{Cabrera1949, Kuramochi2003, Lin2006, Cambel2007}. In our case, the onset of saturation is suppressed, which we attribute to the higher ambient humidity maintaining a larrger population of mobile ions within the meniscus and thereby sustaining the oxidation rate even at elevated fields. The oxide height ($h$) varies with the applied tip–sample voltage ($V_{TS}$)
according to power-law relation $h = k V_{TS}^{n}$, with estimated parameters $k = $ \qty{6.0E-4}{\nm} and $n = $ \num{3.38} (with a coefficient of determination $R^{2} = $ \num{0.985}). The strong correlation confirms field-enhanced oxidation governed by ionic-transport-limited kinetics typical of anodic oxide formation in AlGaAs. Furthermore, the oxide height shows an exponential dependence on Al content, described by $h = h_{\infty}[1 - \exp(-x/x_{0})]$ with $h_{\infty} = $ \qty{21.5}{\nm} and $x_{0} = $ \qty{5.06}{\percent} ($R^{2}=0.75$ owing to limited dataset), indicating that higher Al fractions promote faster oxidation due either to the larger electrochemical potential of Al as compared to Ga or a higher diffusivity of the smaller Al ions, leading to faster oxide growth \cite{Reinhardt1996}. A moderate dependence on ambient humidity was also observed, with increased humidity leading to thicker oxides owing to enhanced ionic mobility within the water meniscus \cite{Kuramochi2003}. The oxide growth rate also scales inversely with tip writing speed, providing flexibility between faster patterning and higher oxide contrast, which is critical for alignment markers suitable for EBL. For the markers used here, typically \qty{15}{\nm} in height and \qty{300}{\nm} in width on Al$_{0.15}$Ga$_{0.85}$As, we employed a tip speed of \qty{0.1}{\um \per \second}, $\mathrm{V_{TS}} = $ \qty{-20}{\volt}, and \qty{60}{\percent} humidity. These conditions yield highly reproducible oxide markers that offer sufficient topographic and optical contrast for alignment during EBL.

Candidate QDs for positioning of CBGs are selected by excluding regions with surface defects, adsorbates (e.g., dust or gold particles), or irregular topography, as these can impact the accuracy of the AFM-NL process. Additionally, QDs with nominal dimensions and shapes are chosen for AFM-NL processing. Around each selected QD, a \qty{10}{\um} $\times$ \qty{10}{\um} square oxide marker is patterned, with each marker requiring \qtyrange{10}{13}{\minute} of writing time. This method is readily scalable for large-area marker arrays by automation, enabling deterministic alignment in subsequent fabrication stages.

\section{Circular Bragg Cavity Fabrication and Device Integration}
\label{sec:CBG_Fab}

Circular Bragg gratings (CBGs) are highly efficient photonic microcavities that enhance light extraction and the radiative emission of centrally embedded quantum emitters, such as QDs, via the Purcell effect. The alternating CBG rings concentrate the optical mode at the cavity center, increasing the density of optical states (LDOS) in the center between the semiconductor and vacuum, thereby shortening the emitter lifetime and boosting collection efficiency of a centrally embedded emitter \cite{Wang2019,Liu2019, MoczalaDusanowska2020, Kolatschek2021, Barbiero2022, Wijitpatima2024}. 

A critical property of CBG-based quantum light sources is their sensitivity to emitter displacement ($\Delta r$) from the cavity center. Even small lateral offsets break the in-plane mode symmetry, coupling the quantum-dot emission preferentially to one of the two orthogonal cavity modes. Further details are discussed in \cref{sec:Pol_PL}. This symmetry breaking results in a measurable polarization imbalance, quantified by the Stokes parameters $S$, which serves as a direct and non-invasive metric of positioning accuracy and its impact on entanglement fidelity \cite{Buchinger2025}. Owing to this sensitivity, polarization-resolved micro-photoluminescence ($\mu$PL) measurements of CBGs provide an ideal framework for studying spatial displacement effects on cavity-emitter coupling and polarization entanglement.

Using AFM-NL markers for deterministic quantum-dot localisation, CBGs with a grating period of $\Lambda = \qty{283}{\nano\meter}$ are fabricated. This period is empirically optimised to align the fundamental cavity mode with the mean emission wavelength of the quantum dots ($\sim\qty{780}{\nano\meter}$).

As established in our recent work \cite{Dhurjati2026a}, the optical performance of circular Bragg gratings (CBGs) is governed by the membrane thickness ($t_M$), trench width ($t_W$), grating period ($\Lambda$), trench depth ($t_d$), and the number of grating rings ($N_r$). Finite-difference time-domain (FDTD) simulations were performed to optimise the device geometry, yielding ideal design parameters of $t_M = \qty{134}{\nano\meter}$, $t_W = \qty{77}{\nano\meter}$, $\Lambda = \qty{276}{\nano\meter}$, $t_d = \qty{83}{\nano\meter}$, and $N_r = 6$.

Due to heterostructure growth constraints, membranes with the ideal thickness of \qty{134}{\nano\meter} are not available. Instead, \qty{150}{\nano\meter}-thick membranes are employed for all fabricated devices. Consequently, the realised CBG structures use $t_M = \qty{150(2)}{\nano\meter}$, $t_W = \qty{92(5)}{\nano\meter}$, $t_d = \qty{105(5)}{\nano\meter}$, and $N_r = 10$. These parameters yield cavities with quality factors of about $Q \approx \num{250}$.

The structures are patterned using electron-beam lithography (EBL) and etched via reactive-ion etching (RIE) in a $Cl_2/Ar$ plasma. The sacrificial Al$_{0.75}$Ga$_{0.25}$As layer beneath the CBG is subsequently removed using \qty{12.5}{\percent} hydrofluoric (HF) acid etching to release the membrane, yielding suspended CBGs with AFM-NL-positioned quantum dots at the cavity center (cf.\ \cref{fig:CBG_Concept}a and \cref{fig:Optical_Characterization}a). The HF etching step is followed by a brief rinse in a mild \qty{0.02}{\percent} potassium hydroxide (KOH) solution to eliminate inorganic fluoride residues. The suspended CBGs are dried in isopropanol using a critical-point dryer (CPD) to prevent capillary-induced collapse.

\section{Optical Characterization Methods}
\label{sec:Opt_Characterization}

The photoluminescence (PL) emission of $\GaAs$ QDs is measured in a low-temperature He-flow cryostat at \qty{4}{\kelvin}. QDs are excited to saturation of the QD neutral exciton ($X^0$), cf. Ref. \cite{Hopfmann2021a}, with a \qty{635}{\nm} continuous-wave laser using a 90:10 beam splitter. Excitation and luminescence collection are conducted with an NIR objective with NA \qty{0.55}. Emitted luminescence is filtered by a \qty{700}{\nm} long-pass filter and resolved by a \qty{1200}{lines/mm} diffraction grating and \qty{50}{\um} slit in a \qty{0.75}{\metre} long spectrometer. This system achieves a spectral resolution of about \qty{35}{\micro \eV}.

Polarization-resolved spectroscopy is performed by introducing a rotating zero-order half-wave plate placed before a fixed linear polarizer in front of the spectrometer. Integrated PL intensities are determined by numerical integration of the spectra over a range of \qty{\pm 5}{\nm} around the $X^0$ QD transition line. This intensity quantity is independent of the QD charge state and is consistent with our previous work \cite{Nie2021, Langer2025a}. The exciton fine-structure is determined by modeling the $X^0$ emission energy as a function of the wave plate angle $\alpha$. The resulting curve is normalized by the mean value and modeled by a sine function, i.e. $\Delta E_{X^0}= a \, \sin(2\alpha+b)$, where $a$ is the amplitude and $b$ is an offset angle. The fine-structure splitting is then given by $\text{FSS} =2a$. See supplemental Sec 3 materials where the modeled data is shown explicitly.

For characterization of the QD-CBG emission polarization characteristics, the Stokes parameters $S_1$ - $S_3$ can be used. They represent the degree of polarization along the linear H/V, diagonal D/A and circular R/L polarization axes and can be determined from the wave plate angle-dependent PL spectra. From these, the principal axis of the emission polarization is determined for each QD independently. This is done by modeling the QD emission intensity $I(\alpha)$ using a sine function $F^I$. Consequently, $\alpha_H$ is defined as $\alpha_H = \alpha(F^I_{max})$ and $\alpha_V = \alpha(F^I_{min})$. This implies that the diagonal (antidiagonal) polarization angle corresponds to $\alpha_D = \alpha_H + 45^\circ$ ($\alpha_A = \alpha_H - 45^\circ$). By utilizing the intensities at the determined angles, i.e. $I_H = I(\alpha_H)$, the Stokes parameters of the QD emission can be calculated by:

\begin{equation}
\label{eq:Stokes_Par}
S_1 = \frac{I_H - I_V}{I_H + I_V}, \quad
S_2 = \frac{I_D - I_A}{I_D + I_A}, \quad
S_3 = \frac{I_R - I_L}{I_R + I_L}\, .
\end{equation}

Note that the $S_3$ cannot be determined using a rotating half-wave plate alone. Due to the symmetry of the investigated QD-CBGs (cf. \cref{sec:Pol_PL}) and detection system no $S_3$-dependence is expected. It is therefore not investigated in this work. Due to the definition $S_1$ along the direction of maximal change, i.e. H/V axis in $I(\alpha)$, and the symmetry of the QD-CBG system, $S_2$ represents effectively a control as no degree of emission polarization is expected. It can therefore be used to determine the accuracy and assumptions of the chosen approach.

\section{Optical Performance}
\label{sec:Opt_Performance}

\begin{figure}[t!]
    \centering
    \includegraphics[width=1\textwidth]{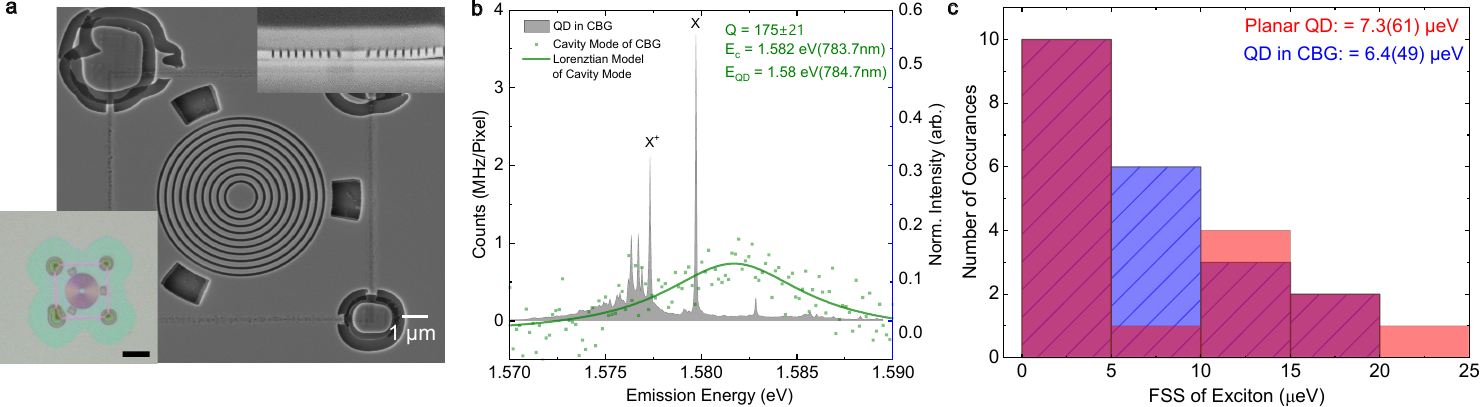}
    \caption{\label{fig:Optical_Characterization}
    Optical characterization of CBGs with embedded QDs.
    (a) SEM image of a positioned QD embedded in a CBG structure, defined relative to the oxide markers and its corners serve as alignment references. The inset shows a focused ion beam (FIB) cross-section of the CBG trench profile. The white scale bar in the inset corresponds to \qty{1}{\micro \metre}. The bottom inset shows an image of free-standing CBG from a confocal microscope, the green-colored region showing the free-standing membrane (scale of \textbf{\qty{5}{\micro \metre}}).
    (b) $\mu$-photoluminescence spectrum of a bright CBG device showing the exciton (X) and trion (X$^+$) emission lines. A Lorentzian fit of the cavity absorption mode is overlaid, yielding a mode energy of \qty{1.582}{\eV} (\qty{783.7}{\nm}) and a quality factor of $Q = $ \num{175(21)} while the exciton emission occurs at \qty{1.580}{\eV} (\qty{784.7}{\nm}). 
    (c) Statistical distribution of exciton fine-structure splitting (FSS) values for planar QDs and QDs in CBGs. Mean values equate to \qty{7.3(61)}{\micro \eV} and \qty{6.4(49)}{\micro \eV} for unprocessed and QDs in CBGs, respectively.}
\end{figure}

High resolution PL spectroscopy is performed on \num{21} positioned CBGs with single QDs and \num{18} QDs in unprocessed pieces of the same wafer material as a control. Clean, well-resolved exciton and trion emission lines were observed in all devices, cf. \cref{fig:Optical_Characterization}b, with the brightest device (CBG~A) exhibiting a \num{245}-fold enhancement in emission intensity compared to average control QDs, corresponding to an integrated intensity of \qty{\approx 75}{\mega\hertz}. 
For CBG-A featuring a $Q = $ \num{175(21)} (RMS bandwidth of \qty{9}{\milli \eV}) and a spectral deviation of \qty{2}{\milli \eV} for the QD exciton emission, cf. \cref{fig:Optical_Characterization}b were measured. SEM and FIB cross-sectional structural analysis of CBGs (\cref{fig:Optical_Characterization}a) shows vertical, smooth CBG sidewalls and well-defined interfaces, consistent with the high optical performance devices.



\section{Emission Polarity}
\label{sec:Pol_PL}

\begin{figure}[H]
    \centering
    \includegraphics[width=1.0\textwidth]{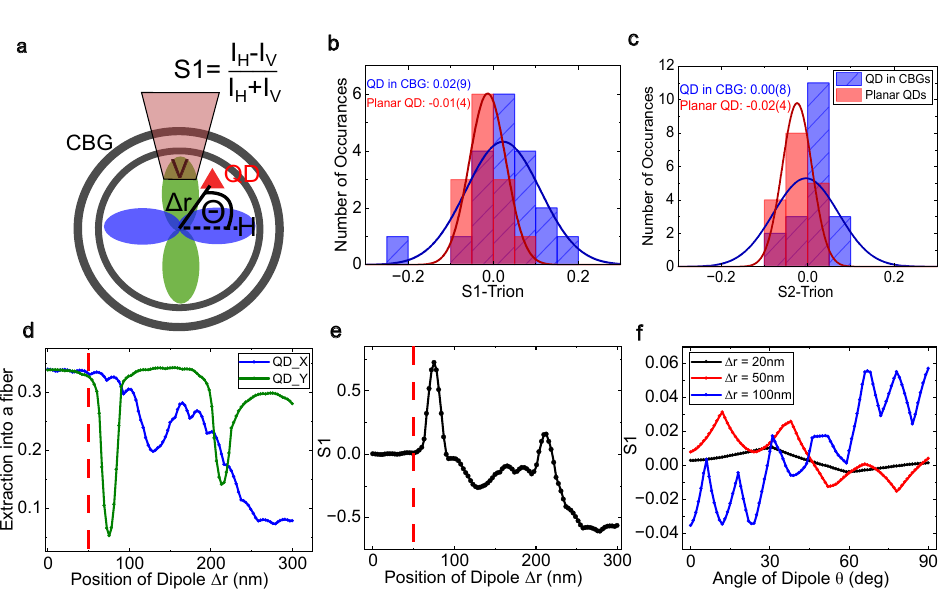}
    \caption{\label{fig:Polarisation_Analysis}
    Polarization analysis of positioned QDs in CBGs using Stokes parameters.
    (a) Schematic illustration of lateral QD displaced from the CBG center by an offset $\Delta r$ and azimuthal angle $\theta$. Spatial profiles of the degenerate linearly H and V polarized fundamental CBG mode are indicated in blue and green, respectively. The emission couples preferentially to one mode depending on $\Delta r$ and $\theta$, leading to a polarization imbalance quantified by $S_1 = (I_H - I_V)/(I_H + I_V)$.
    (b) Statistical distribution of determined $S_1$ values for trion for planar and QD-CBGs, modeled by a normal distributions, with distribution means are denoted.
    (c) Corresponding statistics of $S_2$ for trion.
    (d) Finite-difference time-domain (FDTD) simulations of dipoles oriented along the $x$ and $y$ directions, representing a QD emitter in the CBG. The simulated extraction efficiencies into a collection fiber are plotted as a function of lateral offset $\Delta r$, showing how emitter misalignment modifies the out-coupling efficiency.
    (e) Stokes parameter $S_1$ calculated from the simulated intensities as a function of QD lateral offset $\Delta r$, showing a nearly constant $|S_1| < $ \num{0.05} for $\Delta r < $ \qty{50}{\nm} corresponding to the high entanglement fidelity regime. 
    (f) Simulated variation of $S_1$ with azimuthal angle $\theta$ for different $\Delta r$ values, showing the anisotropic polarization coupling of the CBG mode.}
\end{figure}

As CBGs are highly sensitive to polarization imbalance caused by emitter–cavity displacement, polarization-resolved PL measurements are employed to assess the positioning accuracy achieved by the AFM-NL technique \cite{Buchinger2025}. Polarization-resolved PL measurements yield a neutral exciton ($X^0$) fine-structure splitting (FSS) of \qty{2.97(0.24)}{\micro \eV} for CBG~A, see supplementary Sec 3. The statistical distribution of FSS across all measured CBGs and planar QDs (\cref{fig:Optical_Characterization}d) yields mean values of \qty{7.3(61)}{\micro \eV} and \qty{6.4(49)}{\micro \eV} for QDs and CBGs, respectively. This constitutes a negligible difference and indicates a largely strain-free suspended CBG membrane after processing, which is a crucial requirement for utilizing high-fidelity polarization entangled photon pair sources for quantum communication networks \cite{Yang2022}.

Due to the in-plane rotational symmetry of circular Bragg gratings, the fundamental cavity mode is doubly degenerate and can be expressed as two orthogonal linearly polarized modes, conventionally denoted H and V, as commonly observed in CBG-based quantum dot microcavities \cite{Ates2012, Liu2019}. For the fundamental CBG cavity modes, these modes are spatially organized in the shape of a 4-leaf clover, see illustration in \cref{fig:Polarisation_Analysis}a. The positioning offset between QD and CBG $\Delta r$ therefore modifies how effectively the QD emission couples to the H and V components of the CBG mode. As a consequence, the observable polarization of the extracted QD emission is dependent on $\Delta r$. This effect is also detrimental for the practical use of polarization-entangled photon pair sources. To quantify how much the limited accuracy of the fabricated QD-CBG sources impacts the suitability of these devices for entangled photon pair sources, Stokes parameters from \cref{eq:Stokes_Par} can be used. 


In the absence of mode splitting or strain in the suspended membrane, and under conditions of efficient QD-CBG coupling as evidenced by strong intensity enhancement, see \cref{fig:Optical_Characterization}c, the polarization contrast ($S_1$) is determined primarily by the emitter’s lateral displacement from the cavity center and residual polarization anisotropy of the optical setup \cite{Peniakov2024}. To isolate the positional contribution, Stokes parameters are determined for planar QDs as reference emitters, allowing calibration of the optical detection system’s inherent polarization bias. The QD $X^+$ trion emission line is the primary indicator of the polarized QD-CBG emission characteristic, as it does not exhibit a fine-structure, thereby reflecting only positional asymmetries within the cavity and validating the consistency of our measurements. Across the CBG devices and planar QDs, we have measured the average Stokes parameter values as shown in \cref{tab:StokesStats} (see \cref{fig:Polarisation_Analysis} for trion data and supplementary for exciton data). The extracted mean Stokes parameters for the CBG devices, $S_{1,X^+} = 2(9) \%$, indicate a small but finite polarization contrast compared to planar QDs ($S_1 = -1(4) \%$). The magnitude of this contrast remains low across the dataset, suggesting that residual polarization imbalance originates from limited emitter–cavity misalignment rather than from intrinsic cavity anisotropy or from the optical detection system. 

\begin{table}[t]
\centering
\caption{Average Stokes parameters $S_1$ and $S_2$ for trion ($\mathrm{X^+}$) and exciton ($\mathrm{X^0}$) transitions in CBG-coupled and planar quantum dots. Values are reported in the form $X(Y)$, where $Y$ denotes one standard deviation in the last digits of $X$.}
\label{tab:StokesStats}
\begin{tabular}{lcccc}
\hline
\hline
 & \multicolumn{2}{c}{CBG QDs} & \multicolumn{2}{c}{Planar QDs} \\
\cline{2-5}
Transition & $S_1$ (\%) & $S_2$ (\%) & $S_1$ (\%) & $S_2$ (\%) \\
\hline
$\mathrm{X^+}$ (trion)   & $2(9)$   & $0(8)$   & $-1(4)$ & $-2(4)$ \\
$\mathrm{X^0}$ (exciton) & $1(10)$  & $0(4)$   & $-4(7)$ & $-2(2)$ \\
\hline
\hline
\end{tabular}
\end{table}

To correlate the experimentally observed polarization contrast with the actual spatial displacement of the emitter, FDTD simulations of orthogonal dipoles oriented along the $x$ and $y$ directions within the CBG are performed. These simulations employ the actual measured dimensions of the fabricated CBG structures given earlier in the text. 
The intensity extracted into the collection fiber is evaluated as a function of the dipole’s radial ($\Delta r$) and angular ($\theta$) displacement, cf. \cref{fig:Polarisation_Analysis}d. Contrary to the experimental realization, the simulation assumes that the QD-cavity displacement and dipole axes are aligned, which represents the worst-case scenario regarding the observed polarized emission. Also, due to the ideal symmetry in the simulations, the displacement directions x and y are equivalent. It is therefore more appropriate to employ polar coordinates for the Stokes parameters analysis of the simulated data.

The extracted horizontal and vertical dipole intensities are used to calculate the Stokes parameter $S_1$ (\cref{fig:Polarisation_Analysis}e–f) using \cref{eq:Stokes_Par}. The resulting $S_1(\Delta r)$ dependence confirms that for emitter displacements below $\Delta r < $ \qty{50}{nm}, the polarization contrast remains below \qty{5}{\percent} for all azimuthal orientations, fully consistent with the experimental polarization based CBG measurements.




\section{Positioning Accuracy and Robustness}
\label{sec:Positioning_Accuracy}



To quantitatively relate the observed optical performance and polarization robustness of the fabricated QD–CBG devices to the underlying positioning accuracy achieved by AFM-NL, we analyze the spatial alignment statistics of the positioned structures. 

\begin{figure}[H]
    \centering
    \includegraphics[width=0.6\textwidth]{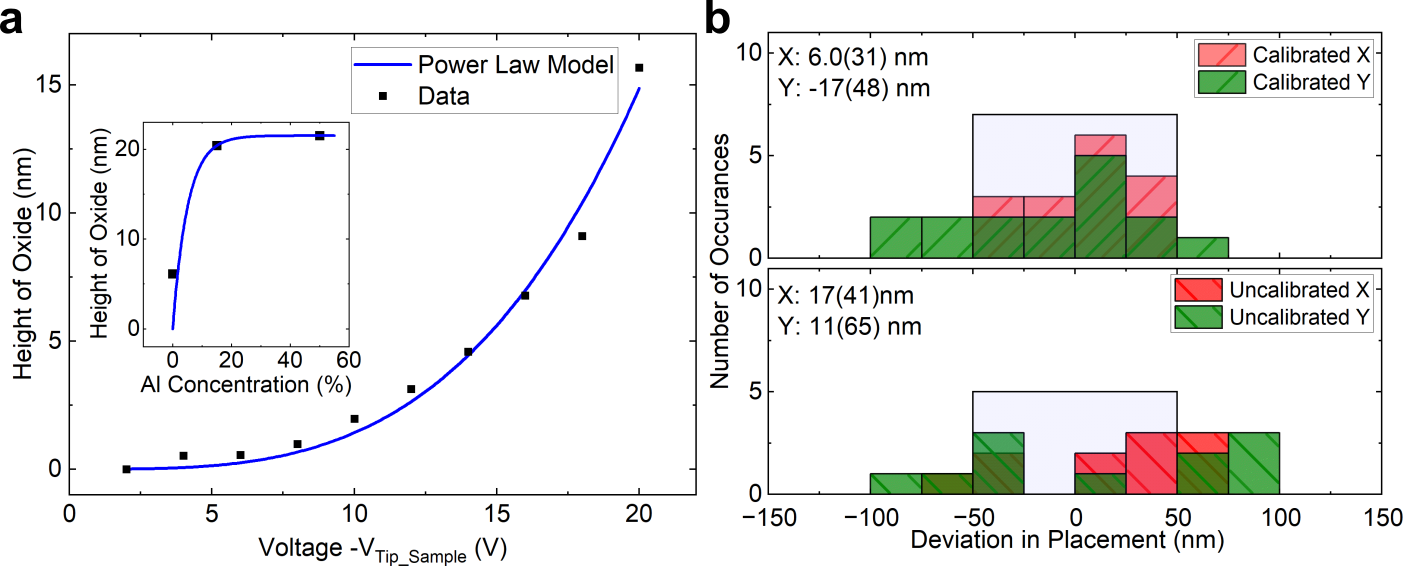}
    \caption{\label{fig:AFM_NL_2}
    (a) AFM-NL oxide height as a function of applied tip–sample voltage, which follows a power-law relation as explained in the text. The inset shows oxide height versus Al concentration at $V_\text{Tip-Sample} = $ \qty{20}{\volt}, modeled by an exponential function. 
    (b) Statistical distribution of QD positioning deviations in the $x$ and $y$ directions obtained from circular modeling of CBG trenches around localized nanoholes, see also Suppl. Sec. \textcolor{red}{XX}. Statistical means and standard deviations were calculated for the calibrated and uncalibrated datasets in $x$ and $y$ axes, respectively. The shaded data corresponds to high entanglement fidelity criteria ($S1< \qty{5}{\percent})$ }
\end{figure}

To determine the achieved positioning accuracy, AFM scans on fabricated QD–CBG devices were performed; however, surface features originating from the QDs are no longer identifiable due to surface modifications introduced during CBG processing. Instead, a reference sample containing unfilled nanoholes (\qtyrange{8}{10}{\nm} deep) at the membrane surface is used to extract the positioning accuracy, as shown in \cref{fig:AFM_NL_2}b, using high-resolution AFM micrographs. Because the lateral dimensions of the nanoholes are relatively large and the etched CBG rings deviate slightly from perfect circularity, the extracted offsets include additional systematic uncertainties arising from the geometric modeling and therefore represent an upper bound on the true positioning deviations. Further details and illustrative examples of the QD–CBG position analysis are provided in the Supplementary Material.

A total of \num{35} test devices are fabricated, and their offsets in the lateral $x$ and $y$ directions between the nanohole centers and the CBG ring centers are determined. Excluding outliers caused by tip crashes, humidity spikes, and tip drag during writing, \num{27} devices are analyzed within a \qty{\pm 100}{\nm} window. An exemplary SEM micrograph of a fabricated QD-CBG device is shown in \cref{fig:Optical_Characterization}(a).

Before applying any spatial axis calibration, the positional error was found to be \qty{17(41)}{\nm}, \qty{11(65)}{\nm} and \qty{74(19)}{\nm} in the $x$, $y$ and radial deviation $\Delta r$, respectively. The larger spread in the $y$-direction is attributed to AFM tip drag during oxide-marker writing, which can shift the effective alignment point away from the marker center depending on tip-sample orientation. To mitigate this effect, an additional AFM scan step to calibrate the alignment of the oxide marker prior to EBL exposure is introduced, resulting in improved $y$-axis reproducibility and reduced spread.

After calibration, the positional offsets improved significantly to \qty{6(31)}{\nm}, \qty{-17(48)}{\nm} and \qty{51(28)}{\nm} in $x$, $y$ and $\Delta r$ respectively . As shown in \cref{sec:Concept}, from the FDTD simulations the brightness of the positioned QD-CBG devices is expected to remain optimal for $\Delta r$ values of up to \qty{50}{\nm}. Within this sub-\qty{50}{\nm} regime, the measured yields reach \qty{100}{\percent} in $x$, \qty{69}{\percent} in $y$, and \qty{63}{\percent} radially. For a more stringent sub-\qty{25}{\nm} criterion, the radial yield is \qty{25}{\percent}. Overall, these yields are comparable to those reported for cryogenic-based positioning techniques \cite{Gschrey2013, Kojima2013, Sapienza2015} and highlight the potential of our room-temperature AFM-based approach for scalable, cost-effective fabrication of high-performance quantum photonic devices.

\section{Conclusions}
\label{sec:Conclusions}
We demonstrate a room-temperature, scalable, and reproducible positioning technique based on atomic force microscopy (AFM)–enabled nano-oxidation lithography (AFM-NL) for the deterministic integration of GaAs quantum dots into CBGs. The method achieves a mean positioning accuracy of \qty{51(28)}{\nano\meter} with a fabrication yield of \qty{63}{\percent}, making it suitable for the realization of high-fidelity entangled photon sources. Optical characterization reveals strong photoluminescence enhancement, reaching up to a factor of 245 for the brightest device compared to planar quantum dots, while polarization-resolved photoluminescence measurements show fine-structure splitting values comparable to planar devices, indicating a strain-free fabrication process.

To directly correlate emitter displacement with optical performance, nanohole--CBG AFM scans are combined with polarization-resolved measurements of the Stokes parameters, which serve as a non-invasive metric for assessing device robustness with respect to polarization entanglement fidelity. For trion emission, we measure average values of $S_{1,\mathrm{X^+}} = \qty{2 \pm 9}{\percent}$ for CBG-coupled quantum dots and $S_{1,\mathrm{X^+}} = \qty{-1 \pm 4}{\percent}$ for planar quantum dots, demonstrating comparable polarization balance. Finite-difference time-domain (FDTD) simulations further show that the polarization contrast $S_1(\Delta r)$ remains below \qty{5}{\percent} for emitter displacements $\Delta r < \qty{50}{\nano\meter}$ across all azimuthal orientations, in excellent agreement with the experimental observations.

Based on the combined positional offset analysis, polarization imbalance measurements, and supporting simulations, we establish AFM-based nano-oxidation lithography as a robust, room-temperature, and scalable platform for the realization of high-brightness quantum light sources suitable for the generation of high-fidelity polarization-entangled photon pairs. Looking forward, full automation of the AFM-NL process is compatible with industrial fabrication workflows. Future directions include extending this technique to other semiconductor material systems, direct fabrication of CBGs using AFM-NL, integration with site-controlled quantum dots defined via AFM-NL, and hybrid AFM-NL / EBL alignment strategies, potentially combined with energy-dispersive X-ray (EDX) analysis for advanced marker-based device integration.

\section*{\label{acknowledgment} Acknowledgment}
We acknowledge Yana Vaynzof for valuable discussions and suggestions. We thank the clean room team of the Leibniz IFW Dresden, especially Ronny Engelhard and Sandra Nestler, for their efforts and expertise in processing of samples. This work was funded by the German federal ministry of research technology and space (BMFTR) projects QR.X, QUARKS, QUIET, and QD-CamNetz (contracts no. 16KISQ016, 16KIS1998K, 16KISQ094, and 16KISQ078).

\section*{\label{das} Data Availability Statement}
All data that support the findings presented in this work are available from the corresponding author upon reasonable request.

\section*{Conflict of Interest}
All authors declare that they have no conflicts of interest.

\clearpage

\section*{Supplementary}

\subsection{Finite Difference Time Domain Simulations}
\label{sec:FDTD_Sim}

To obtain a viable CBG design starting point for the empirical optimization process, device performance simulations are performed. This is achieved by the utilization of a commercial FDTD solver (Lumerical) in  combination with our recently published work on the accurate determination of the cryogenic \AlGaAs refractive index as a function of the emission wavelength \cite{Langer2025b}. To ensure realistic simulation results, the solver was used in full 3D-mode of the whole CBG structure. The key performance indicator used to optimize the CBG design is the fiber-coupled extraction efficiency of a QD-CBG device coupled to a lensed fiber ($NA =$ \num{0.6}), details on this method can be found in our previous works \cite{Nie2021, Langer2025a}. The detailed description and evaluation of the combined simulation and empirical device optimization process is beyond the scope of this work and can be found in Ref. \cite{Dhurjati2026a}.

\begin{figure}[H]
    \centering
    \includegraphics[width=0.6\textwidth]{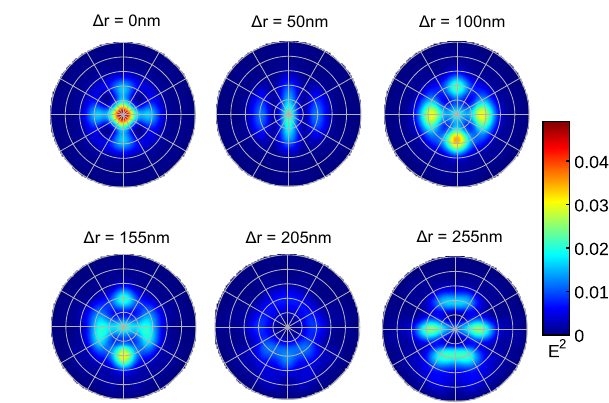}
    \caption{\label{fig:FarField_Emissions_Displacements}
    Simulated far-field emission patterns for QDs at various lateral displacements ($\Delta r$) from the CBG center. 
    Each panel shows the spatial distribution of the electric field intensity ($E^2$) corresponding to the indicated QD displacement values.}
\end{figure}

The extraction efficiency as a function of dipole orientation for various radial displacements, along with the corresponding Stokes vector values for the trion emission, are also presented. The trion mean (standard deviation) values follow the same trend as the exciton, validating the consistency of the optical response discussed in the main text. Furthermore, FDTD simulations of the far-field emission for different radial displacements reveal distinct mode patterns, highlighting the sensitivity of the cavity emission profile to the emitter position(\cref{fig:FarField_Emissions_Displacements}).

\subsection{EDX Spectroscopy}
\label{sec:EDX_Spectroscopy}

\begin{figure}[H]
    \centering
    \includegraphics[width=0.7\textwidth]{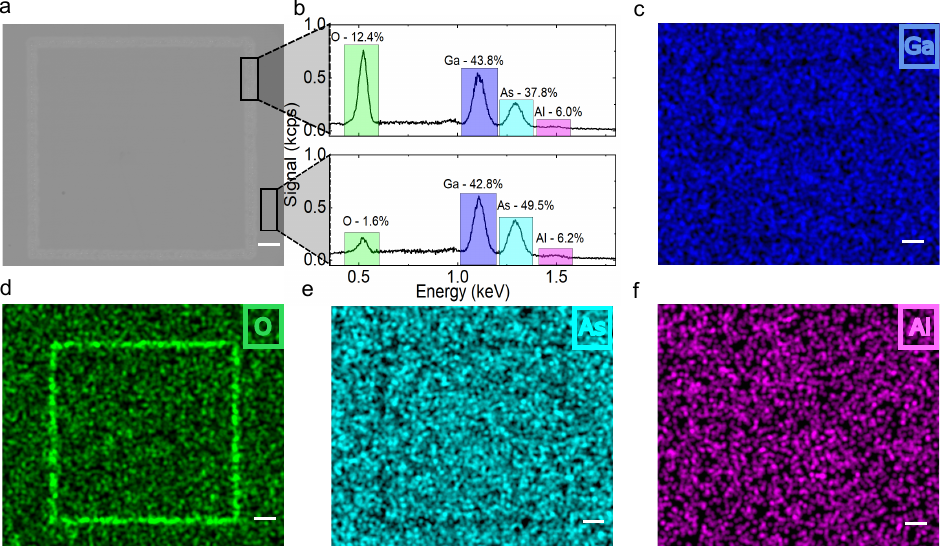}
    \caption{\label{fig:EDX_Scans}
    Scanning electron microscopy (SEM) and energy-dispersive X-ray (EDX) measurements of the AFM nano-oxide marker.
    (a) SEM image of the AFM oxide marker. The white scale bar corresponds to 1~µm. 
    (b) EDX spectra acquired on top of the oxide marker (top) and on the planar surface (bottom), showing characteristic elemental peaks and confirming local oxidation.
    (c)--(f) EDX elemental maps of Ga, O, As, and Al, respectively, showing the presence of oxygen in the oxide marker, suppression of As, and a mild Ga signal. 
    }
\end{figure}

Energy-dispersive X-ray spectroscopy (EDX) of the AFM nano-oxide marker probes the local composition by detecting the characteristic X-rays emitted after inelastic excitation of atoms by the incident electron beam. At the accelerating voltage used here, the information depth is on the order of a few tens of nanometres, so the signal arises predominantly from the oxide and the membrane surface rather than the bulk. Limited-area spectra acquired at the positions marked by black boxes in \cref{fig:EDX_Scans}a, one on the oxide marker and one on the adjacent membrane show a pronounced oxygen peak on the marker compared to the membrane surface (\cref{fig:EDX_Scans}b).
Elemental maps over the full field of view (\cref{fig:EDX_Scans}c–f) reveal an oxygen-rich feature that reproduces the marker outline seen in SEM, a suppression of the arsenic signal, and a reduced yet still present gallium signal. Aluminium and Gallium appear broadly distributed across the surface, consistent with their presence in both the oxide and the underlying AlGaAs. The diminished As signal within the marker likely reflects the preferential formation of Al–O and Ga–O species and the volatilization of unstable As oxides, rather than a mapping artifact consistent with prior reports that AFM-induced oxides on GaAs/AlGaAs are dominated by stable Al and Ga oxide phases, with As oxides being transient and difficult to detect by EDX.

\subsection{Nanohole-CBG Positioning}

\begin{figure}[H]
    \centering
    \includegraphics[width=0.6\textwidth]{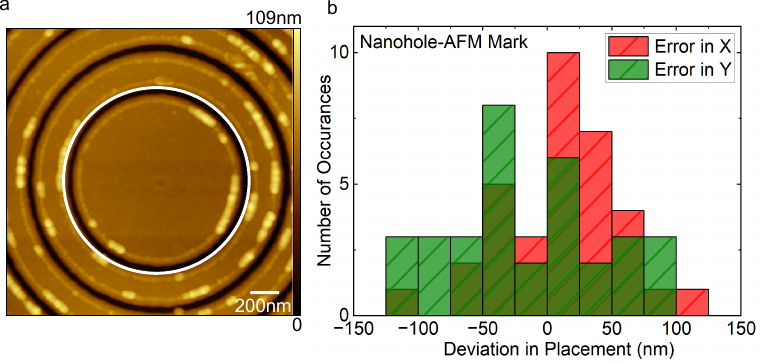}
    \caption{\label{fig:Nanohole_CBG_Positioning}
    Position analysis of nanohole--CBG alignment.
    (a) AFM image of a CBG containing a nanohole showing the inner rings. The nanohole remains clearly visible despite the large height range and post-processing due to which we have used this for positioning analysis. The white circle indicates a circular fit used to estimate the positional offset between the CBG center and the nanohole center. 
    (b) Statistical distribution of positioning offsets in the $x$ and $y$ directions obtained from circular fitting of CBG trenches around localized nanoholes. The full dataset is shown for both $x$ and $y$ directions. 
    }
\end{figure}

The \cref{fig:Nanohole_CBG_Positioning}a shows an AFM scan of the CBG with a nanohole centered using AFM markers where the nanohole is clearly visible despite the fabrication imperfections. The CBG trenches are fit using imageJ software to calculate the center and was averaged across various CBG trenches to limit the error from trench imperfections. The AFM-based determination of the nanohole center has an estimated positional uncertainty of $\approx$ 7 nm (rms radial), dominated by trench-fit and morphology errors; the pixel-sampling contribution (1.8 nm $px^{-1}$) is below 0.3 nm and therefore negligible. From the full datasets, the mean offsets are \(15(57)\,\mathrm{nm}\), \(-11(79)\,\mathrm{nm}\), and \(89(37)\,\mathrm{nm}\) for the \(x\), \(y\), and radial directions, respectively for the uncalibrated method and \(1.7(54)\,\mathrm{nm}\), \(-37(59)\,\mathrm{nm}\), and \(72(48)\,\mathrm{nm}\) for the calibrated method. After excluding outliers and restricting to \(|x|,|y|\le100~\mathrm{nm}\), the results are summarized in Table~\ref{tab:calibration}. Calibration reduces the radial deviation $\Delta r$ from $74(19)$~nm to $51(28)$~nm, improving the sub-50~nm yield from $9.1\%$ to $63\%$ demonstrating a substantial enhancement in positional accuracy and reproducibility of the AFM-assisted alignment.

\begin{table}[H]
\centering
\caption{Mean offsets $\mu$ and standard deviations $\sigma$ (nm) for the uncalibrated and calibrated datasets.
Yields denote the percentage of devices within 25~nm and 50~nm subsets.}
\small
\begin{tabular}{lccc|cc}
\toprule
 & $\mu_x(\sigma_x)$ & $\mu_y(\sigma_y)$ & $\mu_{\Delta r}(\sigma_{\Delta r})$ & sub-25 (\%) & sub-50 (\%) \\
\midrule
Uncalibrated & $17(41)$ & $11(65)$ & $74(19)$ & $x$:18, $y$:9.1, $\Delta r$:0 & $x$:64, $y$:36, $\Delta r$:9.1 \\
Calibrated   & $6.0(31)$ & $-17(48)$ & $51(28)$ & $x$:56, $y$:44, $\Delta r$:25 & $x$:100, $y$:69, $\Delta r$:63 \\
\bottomrule
\end{tabular}
\label{tab:calibration}
\end{table}

\subsection{Polarisation Characterisation}

\begin{figure}[t]
    \centering
    \includegraphics[width=0.7\textwidth]{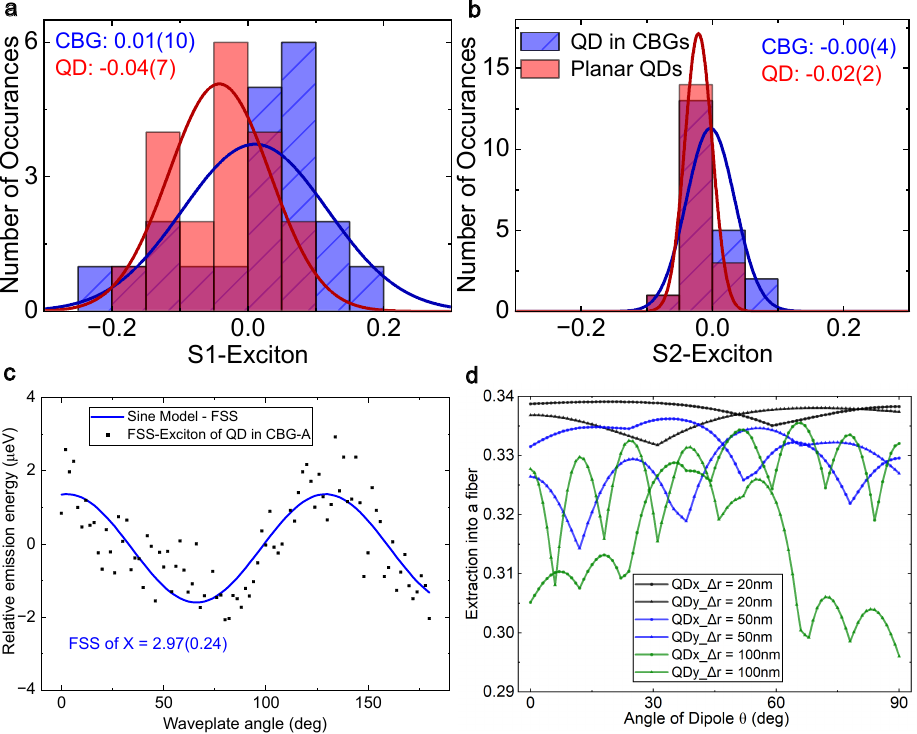}
    \caption{\label{fig:Pol_CBG}
    Optical characterization and polarization analysis of positioned QDs in CBGs using Stokes parameters.
    (a) Statistical distribution of determined $S_1$ values for exciton for planar and QD-CBGs, modeled by a normal distributions, with distribution means are denoted.
    (b) Corresponding statistics of $S_2$ for exciton.
    (b) Simulated extraction efficiency into a collection fiber for $x$ and $y$ oriented dipoles in a CBG as a function of dipole angle ($\theta$) for various radial displacements ($\Delta r$).
    Simulations were performed from $0$ to $90^{\circ}$, as the CBG is symmetric in the other three quadrants. 
    (c)--(d) Stokes parameters $S_1$ and $S_2$ extracted from polarization-resolved PL measurements of the trion emission for planar QDs and QDs embedded in CBGs.}
\end{figure}

Polarisation-resolved measurements of the exciton (X) transition for both CBG-coupled and planar quantum dots are presented in \cref{fig:Pol_CBG}a–b. The Stokes parameters $S_1$ and $S_2$ were extracted as described in the main text. For the exciton transition, the measured values of $S_1$ are $1(10)$\% for the CBG device and $-4(7)$\% for the planar reference QD. The corresponding values of $S_2$ are $0(4)$\% and $-2(2)$\%, respectively. 

The fine-structure splitting (FSS) of the brightest device (CBG-A) is shown in \cref{fig:Pol_CBG}c. A value of $\mathrm{FSS} = 2.97 (0.24)~\mu\mathrm{eV}$ is extracted for the exciton transition, confirming the preservation of near-degenerate exciton states in the CBG architecture.

To further understand the robustness against misalignment and dipole orientation, finite-difference time-domain (FDTD) simulations were performed. The extraction efficiency and the resulting Stokes parameters were calculated as a function of radial displacement $\Delta r$ and dipole orientation angle $\theta$. The angular dependence of extraction efficiency for several radial offsets $\Delta r$ is presented in \cref{fig:Pol_CBG}d.

\clearpage

\printbibliography

\end{document}